# Spin exchange dynamics in 4H SiC monocrystals with different nitrogen donor concentrations

M. Holiatkina[1], A. Pöppl[2], E. Kalabukhova[3], J. Lančok[4], D. Savchenko[1,4]

[1]*Department of General Physics and Modeling of Physical Processes, National Technical University of Ukraine "Igor Sikorsky Kyiv Polytechnic Institute", pr. Beresteiskyi 37, Kyiv, 03056, Ukraine*

[2]*Felix Bloch Institute for Solid State Physics, University of Leipzig, Linnestrasse. 5, Leipzig, 04103, Germany*

[3]*Department of optics and spectroscopy, V.E. Lashkaryov Institute of Semiconductor Physics NAS of Ukraine, pr. Nauky 41, Kyiv, 03028, Ukraine*

[4]*Department of Analysis of Functional Materials, Institute of Physics of the CAS, Na Slovance 2, 18200, Prague 8, Czech Republic*

*Corresponding author e-mail: dariyasavchenko@gmail.com

## Abstract

4H silicon carbide (SiC) polytype is preferred over other SiC polytypes for high-power, high-voltage, and high-frequency applications due to its superior electrical, thermal, and structural characteristics. In this manuscript, we provide a comprehensive study of the spin coupling dynamic between conduction electrons and nitrogen (N) donors in monocrystalline 4H SiC with various concentrations of uncompensated N donors from $10^{17}$ cm$^{-3}$ to $5\times10^{19}$ cm$^{-3}$ by continuous wave, pulsed EPR, and microwave perturbation techniques at $T = 4.2$-$300$ K. At low temperatures two triplets due to N donors in cubic ($N_k$) hexagonal ($N_h$) positions and triplet arisen from spin-interaction between $N_h$ and $N_k$ were observed in 4H SiC having $N_d - N_a \approx 10^{17}$ cm$^{-3}$. A single S-line ($S = 1/2$) dominates the EPR spectra in all investigated 4H SiC monocrystals at high temperatures. It was established that this line occurs due to the exchange coupling of localized electrons (dominate at low temperatures) and non-localized electrons (dominate at high temperatures). The localized electrons were attributed to $N_h$ for $N_d - N_a \approx 10^{17}$ cm$^{-3}$ and $N_k$ donors for $N_d - N_a \geq 5 \times 10^{18}$ cm$^{-3}$. We have concluded that the conduction electrons in 4H SiC monocrystals are characterized by $g_\parallel = 2.0053(3)$ $g_\perp = 2.0011(3)$ for $N_d - N_a \leq 5 \times 10^{18}$ cm$^{-3}$ and $g_\parallel = 2.0057(3)$ and $g_\perp = 2.0019(3)$ for $N_d - N_a \approx 5 \times 10^{19}$ cm$^{-3}$. Using the theoretical fitting of the temperature variation of S-line EPR linewidth in 4H SiC having $N_d - N_a \leq 5 \times 10^{18}$ cm$^{-3}$, the energy levels of 57-65 meV that correlate with the valley-orbit splitting values for $N_k$ donors in 4H SiC monocrystals were obtained.

**Keywords:** silicon carbide, EPR, spin exchange, nitrogen donors

**Introduction**

Silicon carbide (SiC) stands for a wide-bandgap semiconductor compound consisting of silicon (Si) and carbon (C), with a high melting point (~2730°C) and high hardness (~9.5 on the Mohs scale). It possesses such unique key properties as high thermal conductivity, mechanical strength, exceptional chemical resistance, low coefficient of thermal expansion, and high radiation resistance.[1] All these features make this material appropriate for high-temperature applications, aerospace and automotive industries, harsh environments, electronic devices, high-frequency devices, nuclear power plants, space exploration, photonics, and biosensors.[2-13] Recently, it was also reported that SiC is a promising material as a novel dark matter detector.[14]

There are over 250 known polytypes of SiC,[15] and the most common ones are: cubic (3C polytype) with a zincblende and hexagonal (4H and 6H polytypes) with wurtzite crystal structures. Different SiC polytypes have different stacking sequences determining crystal symmetry and physical properties. If one considers the C atom positions within a SiC bilayer so that they build a hexagonal structure ("A" site) and then the following bilayer can position its C atom on the "B"/"C" lattice sites, and as a result, each of bilayers in SiC can be oriented in three possible arrangements only regarding the lattice whereas the tetrahedral bonding is maintained. For example, 3C SiC possesses a stacking sequence ABCABC... while for 4H SiC, it is ABCB.

The hexagonal 4H polytype of SiC is preferred over 3C and 6H polytypes for applications in high-power, high-voltage, and high-frequency devices owing to its exceptional electrical, thermal, and structural properties such as higher values of breakdown voltage, thermal conductivity, electron mobility, crystal quality, and broader bandgap.[16-18]

Nitrogen (N) has a shallow donor energy level in the SiC bandgap, and it is a main donor impurity in SiC, which can give an extra electron to the conduction band, contributing to its n-type conductivity. For example, in 4H SiC epilayers with uncompensated N donor concentration $N_d - N_a$ from $3 \times 10^{15}$ cm$^{-3}$ to $2 \times 10^{16}$ cm$^{-3}$, the activation energy values were obtained as 40-65 meV for the N donors at hexagonal ("h"), $N_h$, and 105-125 meV for the N donors at cubic ("k"), $N_k$, sites using admittance spectroscopy and Hall effect measurements.[18,19] The low ionization energy of N donors means that they can be easily ionized at low temperatures, forming free electrons in SiC. The N donor concentration significantly impacts the SiC electronic and optical properties. Controlling the N donor concentration makes it possible to fabricate the SiC material with preferred characteristics for particular device applications such as high-temperature and high-power electronics, photovoltaics, and optoelectronics.

The N-doped 4H SiC has wide applications in power electronic devices (Schottky diodes, power MOSFETs, and bipolar transistors), light-emitting devices (high-power and high-brightness LEDs, laser diodes), photovoltaic devices, radiation detectors, high-temperature electronics, and sensors.[20-28] As a result, N-doped 4H-SiC is a promising candidate for next-generation devices that require high performance and reliability. Therefore, studying the magnetic and electronic properties of 4H SiC with different N donor concentrations can help researchers understand the N doping mechanisms and optimize the N-doping process for specific material applications.

Electron paramagnetic resonance (EPR) spectroscopy, both in continuous wave and pulse regimes, is a powerful, effective, and non-destructive method for studying the magnetic and electronic features of N donors in SiC.

According to Refs. 29-31, the continuous wave and pulsed EPR spectra measured at $T = 4.2$-40 K of 4H SiC wafers with $N_d - N_a < 10^{18}$ cm$^{-3}$ are characterized by three triplets owing to hyperfine coupling with $^{14}$N nuclei ($I = 1$, 100 % nat. ab.): the line triplet due to $N_k$, a line triplet due to $N_h$, and a triplet $N_x$ due to spin-coupling between $N_h$ and $N_k$. As it follows from Fig. 2 in Ref. 30, at $T > 50$ K, the $N_h$ and $N_x$ triplets disappear in the EPR spectrum of 4H SiC, and a broad line emerges, but no detailed investigation of the temperature dependence and parameters of spin Hamiltonian for this paramagnetic center was reported.

In the 4H SiC wafers with $N_d - N_a$ from $10^{18}$ to $10^{19}$ cm$^{-3}$, the continuous wave EPR spectra at $T = 4.2$ K showed a broad and intense line with no hyperfine structure and the $N_k$ triplet of very low intensity.[30] The appearance of this intense single line was interpreted by concluding that the position of the Fermi level is fixed at the donor (+/0) energy level of aggregates of 2 (or more) $N_k$ donor centers between the isolated $N_k$ donor energy levels. However, no temperature-dependent study of this paramagnetic center was performed. In 4H SiC wafers having $N_d - N_a \approx 10^{19}$ cm$^{-3}$, the asymmetric Dysonian EPR line owing to exchange interaction between localized and conduction electrons, was detected at $T = 7$-140 K.[32]

Thus, the nature of the single EPR lines that appeared in 4H SiC monocrystals with various N donor concentrations was not determined concerning its material electronic and magnetic properties. In this paper, we have investigated electron and magnetic properties of the S-line that appeared due to exchange-coupled localized and non-localized electrons in 4H SiC monocrystals having various N donor concentrations over a broad temperature range utilizing contactless microwave (MW) perturbation measurements and EPR spectroscopy in continuous wave and pulse regimes.

**Materials and methods**

Monocrystalline 4H SiC with a concentration of uncompensated N donors of $N_d - N_a \approx 10^{17}$ cm$^{-3}$ were grown by sublimation sandwich method,[33] and the monocrystalline 4H SiC with $N_d - N_a \approx 5 \times 10^{18} \div 5 \times 10^{19}$ cm$^{-3}$ were grown by the modified Lely method.[34]

X-band ($v_0 \sim 9.4$ GHz) continuous wave EPR in monocrystalline 4H SiC with $N_d - N_a \approx 5 \times 10^{18} \div 5 \times 10^{19}$ cm$^{-3}$, MW perturbation measurements in monocrystalline 4H SiC with $N_d - N_a \approx 10^{17} \div 5 \times 10^{19}$ cm$^{-3}$ and two-pulse field-sweep electron spin echo (FS ESE) spectra in 4H SiC with $N_d - N_a \approx 10^{17}$ cm$^{-3}$ were performed on the Bruker ELEXSYS E580 spectrometer. Continuous wave EPR and MW perturbation experiments were performed in the temperature range from 4.2 K to 300 K with Bruker ER 4122 SHQE SuperX High-Q cylindrical TE$_{011}$ cavity and variable temperature helium-flow cryostat ER 4112HV. We used the following experimental parameter: MW power level was set to 0.4743 mW, modulation frequency was 100 kHz, modulation amplitude was set to 0.5 -1.0 mT (depending on the EPR linewidth), conversion time was 70 ms, and the spectral resolution was selected as 2048 points. The 4H SiC samples were placed on a fused quartz rod with a diameter of 4 mm. The *1,1-diphenyl-2-picrylhydrazyl* free radical ($g = 2.0036$) was used as a reference sample. The quality factor ($Q$-factor) value for the unloaded and loaded cavity was measured when the MW power level was set to 0.07518 mW (33 dB).

X-band ($v_0 \sim 9.6$ GHz) continuous wave EPR in monocrystalline 4H SiC with $N_d - N_a \approx 10^{17}$ cm$^{-3}$ were performed on Bruker EMX EPR spectrometer equipped with ESR 900 He cryostat from Oxford Instruments. The continuous wave EPR spectra were measured using ER 4105DR double rectangular cavity operating in the TE104 mode using the ultramarine reference sample as an intensity standard. The experimental parameters were the following: MW power level = 0.1517 mW, spectral resolution = 2048 points, modulation frequency = 100 kHz, modulation amplitude = 0.2 mT, time constant = 40.96 ms, and conversion time = 120 ms. For MW perturbation experiments the super-high-Q cavity was used. The $Q$-factor value was measured when the MW power level was set to 0.6325 mW (25 dB).

The temperature variation of FS ESE spectra was studied utilizing EN 4118X-MD5 cavity equipped with ER 4118CF cryostat. The FS ESE spectra were obtained using a two-pulse Hahn echo sequence: $\pi/2 - \tau - \pi - \tau -$ echo with the pulse lengths: $\pi/2 = 100$ ns, $\tau = 1200$ ns, $\pi = 200$ ns.

The EPR spectra parameters of the paramagnetic centers in monocrystalline 4H SiC were analyzed utilizing the following spin-Hamiltonian:

$$\hat{H} = H_{EZI} + H_{NZI} + H_{HFI} + H_{ZFS} + H_{QI} = \mu_B \mathbf{B g S}/\hbar - g_n \mu_n \mathbf{B I}/\hbar + \sum_i \mathbf{S A}_i \mathbf{I}_i + \mathbf{S D S} + \mathbf{I P I} \qquad (1)$$

where $H_{EZI}$ and $H_{NZI}$ describe electron and nuclear Zeeman interactions, $H_{HFI}$ describes hyperfine interactions (for $I > 0$), $H_{ZFS}$ describes zero-field splitting (for $S \geq 1$), $H_{QI}$ describes quadrupole interaction (for $I \geq 1$), $\mu_B$ – Bohr magneton; $\mathbf{B} = (0, 0, B_0)$ – externally applied magnetic field; $B_0$ – is the resonance magnetic field position, $\mathbf{g}$ – is the electron $g$-tensor; $\mathbf{S}$ – is the electron spin operator; $\hbar = h/2\pi$ – is the Plank constant; $g_n$ – is the nuclear $g$-factor; $\mu_n$ – nuclear magneton; $\mathbf{I}_i$ – is the nuclear spin operator of $i$-th nucleus; $\mathbf{A}_i$ – tensor of hyperfine interaction of $i$-th nucleus; $\mathbf{D}$ – zero-field splitting tensor, $\mathbf{P}$ – nuclear quadrupole interaction tensor.

For cubic system with $n$-order symmetry axis ($n \geq 3$) at $\mathbf{B} \| [001]$, the symmetry of $\mathbf{g}$-tensor is axial and $g_1 = g_\|$, a $g_2 = g_3 = g_\perp$ and the corresponding hyperfine interaction tensor should be written as: $A_\| = a_l + 2b_l$ and $A_\perp = a_l - b_l$ ($a_l$ – isotropic hyperfine interaction constant of $l$-th atom, $b_l$ – anisotropic hyperfine interaction constant of $l$-th atom).

For the paramagnetic center with $S = 1/2$ and $I = 0$, only $H_{EZI}$ and $H_{NZI}$ from Eq. (1) should be considered, and a single line in the EPR spectrum will be detected. For the paramagnetic center described by $S = 1/2$ and $I = 1$, the terms $H_{EZI}$, $H_{NZI}$, $H_{HFI}$, and $H_{QI}$ from Eq. (1) should be taken into account, and a triplet in the EPR spectrum is observed ($2I + 1 = 3$ lines). For the paramagnetic center having $S = 1$ and $I = 1$, all the terms from spin Hamiltonian in Eq. (1) are considered, and two triplets in the EPR spectrum will be observed.

The first derivative of asymmetric Dysonian lineshape characterized by the dispersion and absorption components was fitted with experimental spectra by using the Matlab R2017a software package (MathWorks ®, Natick, USA) with the following expression assuming the skin layer is formed on the flat plate of thickness $2d$:[35,36]

$$\frac{dI}{dB} = A\frac{2x}{\left(1+x^2\right)^2} + D\frac{1-x^2}{\left(1+x^2\right)^2}, \qquad (2)$$

$$A = \frac{\sinh p + \sin p}{2p(\cosh p + \cos p)} + \frac{1 + \cosh \cos p}{(\cosh p + \cos p)^2}, \quad D = \frac{\sinh p - \sin p}{2p(\cosh p + \cos p)} + \frac{\sinh p \sin p}{(\cosh p + \cos p)^2}, \qquad (3)$$

where $x = 2(B - B_0)/\sqrt{3}\Delta B_{pp}^L$, $\Delta B_{pp}^L$ – is the peak-to-peak Lorentzian linewidth, $p = 2d/\delta$, $\delta = \sqrt{2/\mu_0 \nu_L \sigma_{ac}}$ – is the skin layer thickness, $\nu_L$ – is the resonance frequency value for the sample-loaded cavity, $\mu_0$ – is the vacuum magnetic permeability, $\sigma_{ac}$ – intrinsic ac conductivity. The $A$ and $D$ coefficients are related to the Dysonian line asymmetry ratio $A/B$ as $A/B = 1 + 1.5 \times D/A$, which is proportional to ac conductivity.[37-39] The case when $D/A = 0$, $A/B = 1$ corresponds to the symmetrical Lorentzian lineshape.

For the simulation of EPR spectra from localized N donors, the 'pepper' function from the Easyspin 5.2.28 toolbox[40] using spin Hamiltonian from Eq. (1) was utilized. The temperature variation of spin susceptibility and EPR linewidth fitting with theory was performed in OriginPro 8 (OriginLab, USA).

**Experimental results**

A. Microwave perturbation measurements

The microwave perturbation technique has been used at $T = 4.2\text{-}300$ K to obtain the activation energies for N donors in the investigated samples. The cavity was filled with the investigated sample having a smaller volume than the cavity, and the shifts in cavity Q-factor and MW frequency values were detected. As a result, from the temperature variation of microwave losses, $\Delta = Q_L^{-1} + Q_0^{-1}$ ($Q_L$ and $Q_0$ – denote the $Q$-factor for the loaded and unloaded cavity, correspondingly) and frequency shift $\delta = (\nu_0 - \nu_L)/\nu_0$ ($\nu_0$ – is the resonance frequency value for the unloaded cavity) in the investigated 4H SiC samples having different N donor concentrations, we have estimated the temperature change in conductivity ($\sigma$) as:[41-44]

$$\sigma = 2 \times \pi \times \nu_L \times \varepsilon_0 \frac{\frac{\alpha}{N^2} \times \frac{\Delta}{2}}{\left(\frac{\alpha}{N} - \delta\right)^2 + \left(\frac{\Delta}{2}\right)^2} \quad (4)$$

where $\varepsilon_0$ is the vacuum dielectric permittivity, $\alpha$ – refers to the filling factor (for TE$_{011}$ cavity $\alpha = 2V_S/V_C$, $V_S$ – is the sample volume, $V_C$ – is the cavity volume), and $N = -\alpha \times \nu_L / \delta$ is the depolarization factor.

The plot of the natural logarithm of $\sigma$ dependence in the coordinates of $1000/T$ obtained from Eq. (4) for monocrystalline 4H SiC samples having various N donor concentrations is represented in Fig. 1.

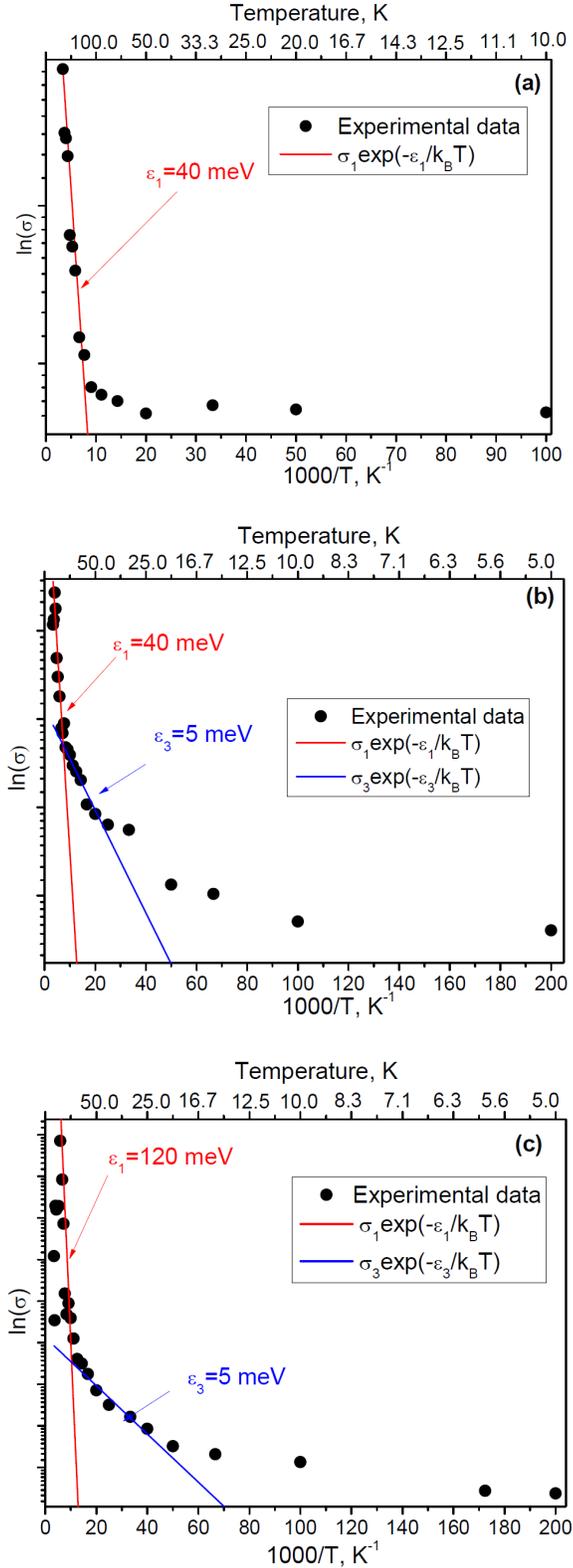

FIG. 1. Natural logarithm of $\sigma$ vs. $1000/T$ derived from Eq. (1) (black dots) for monocrystalline 4H SiC with $N_d - N_a \approx 10^{17}$ cm$^{-3}$ (a), $N_d - N_a \approx 5 \times 10^{18}$ cm$^{-3}$ (b), $N_d - N_a \approx 5 \times 10^{19}$ cm$^{-3}$ (c). Red and blue solid lines show the output from the fitting with Eq. (5).

In the general case, the conductivity in semiconductors of n-type can be expressed by the sum of three contributions – thermally activated carrier density in the conduction band having an activation energy of $\varepsilon_1$, and nearest-neighbor hopping processes of electrons from neutral donors to neutral donors ($2D^0 \rightarrow D^- + D^+$) with activation energy $\varepsilon_2$ and from neutral donors to empty positively charged donors with activation energy $\varepsilon_3$:[45-48]

$$\sigma(T) = \sigma_1 \exp(-\varepsilon_1/k_B T) + \sigma_2 \exp(-\varepsilon_2/k_B T) + \sigma_3 \exp(-\varepsilon_3/k_B T), \quad (5)$$

where $k_B$ – is the Boltzmann constant.

From Fig. 1, it follows that in monocrystalline 4H SiC samples with $N_d - N_a \approx 5 \times 10^{19}$ cm$^{-3}$ at $T = 297\text{-}170$ K and in 4H SiC samples with $N_d - N_a \approx 5 \times 10^{18}$ cm$^{-3}$ at $T = 297\text{-}250$ K the conductivity grows owing to conduction electron scattering by ionized N donors.

The electron transition process from N donor energy levels to the conduction band takes place at $T = 297\text{-}90$ K for 4H SiC with $N_d - N_a \approx 10^{17}$ cm$^{-3}$, at $T = 250\text{-}130$ K for 4H SiC with $N_d - N_a \approx 5 \times 10^{18}$ cm$^{-3}$ and from 170 K to 90 K for 4H SiC having $N_d - N_a \approx 5 \times 10^{19}$ cm$^{-3}$.

The hopping process of electrons between N donors appears at $T < 130$ K in monocrystalline 4H SiC samples $N_d - N_a \approx 5 \times 10^{18}$ cm$^{-3}$ and at $T < 90$ K in 4H SiC monocrystals with $(N_D - N_A) \sim 5 \times 10^{19}$ cm$^{-3}$, whereas this process is absent for 4H SiC with $(N_D - N_A) \sim 10^{17}$ cm$^{-3}$. Moreover, a variable-range hopping process at very low temperatures can be expected in the 4H SiC having $N_d - N_a \geq 5 \times 10^{18}$ cm$^{-3}$.

Based on the fitting of terms in Eq. (5) with experimental data represented in Fig. 1, we have obtained that $\varepsilon_1 = 40$ meV for 4H SiC with $N_d - N_a \approx 10^{17} \div 5 \times 10^{18}$ cm$^{-3}$ and $\varepsilon_1 = 120$ meV for 4H SiC with $N_d - N_a \approx 5 \times 10^{19}$ cm$^{-3}$. The $\varepsilon_3$ value was estimated as 5 meV in 4H SiC with $N_d - N_a \approx 5 \times 10^{18} \div 5 \times 10^{19}$ cm$^{-3}$. From this $\varepsilon_3$ value and knowing the Bohr radius for 4H SiC $a_B = 1.2$ nm,[49] one can determine the density of states at the Fermi energy following Ref. [50] for monocrystalline 4H SiC having $N_d - N_a \approx 5 \times 10^{18} \div 5 \times 10^{19}$ cm$^{-3}$ as $N(E_F) \approx \left( \varepsilon_3 \times a_B^3 \times 4\pi/3 \right)^{-1} = 2.8 \times 10^{22}$ eV$^{-1} \times$cm$^{-3}$.

B. EPR measurements

Fig. 2 illustrates the low-temperature EPR spectra measured at $T = 10$ K in monocrystalline 4H SiC having different N donor concentrations at **B**||**c**.

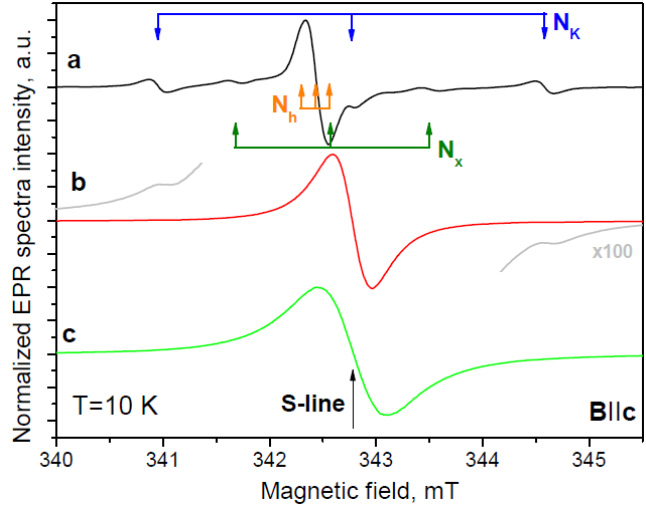

FIG. 2. EPR spectra recorded in monocrystalline 4H SiC having $N_d - N_a \approx 10^{17}$ cm$^{-3}$ (a), $N_d - N_a \approx 5 \times 10^{18}$ cm$^{-3}$ (b), $N_d - N_a \approx 5 \times 10^{19}$ cm$^{-3}$ (c). $T = 10$ K, **B**||**c**. The magnetic field value was adjusted to the same MW frequency value of $v_0 \sim 9.616784$ GHz. The intensity of EPR spectra was normalized to its maximum magnitude.

From the EPR spectra analysis, it was found that the EPR spectra in 4H SiC monocrystals with $N_d - N_a \approx 10^{17}$ cm$^{-3}$ consist of three triplets owing to the hyperfine coupling of the unpaired electron with $^{14}$N nuclei ($I = 1$): the line triplet due to N$_k$ ($S = 1/2$, $g_\parallel = 2.0043(3)$, $g_\perp = 2.0013(3)$, $A_{\parallel}$ = 50.95 MHz, $A_{\perp}$ = 51 MHz),[51] a line triplet due to $N_h$ ($S$ = 1/2, $g_{\parallel}$ = 2.0063(3), $g_{\perp}$ = 2.0005(3), $A_{\parallel}$ = 2.9 MHz, $A_{\perp}$ = 2.7 MHz), and a triplet $N_x$ ($S$ = 1, $g_{\parallel}$ = 2.0053(3), $g_{\perp}$ = 2.0010(3), $A_{\parallel}$ = $A_{\perp}$ = 25.6 MHz, $D$ = 1.7-14 MHz) caused by spin-interaction between $N_h$ and $N_k$.[29-31]

The monocrystalline 4H SiC having $N_d - N_a \approx 5 \times 10^{18}$ cm$^{-3}$ (Fig. 2b) at $T$ = 10 K revealed a broad, intense line characterized by $S$ = 1/2, $g_{\parallel}$ = 2.0045(3), $g_{\perp}$ = 2.0010(3) labeled as S-line having no hyperfine structure along with the $N_k$ triplet of very low intensity. In 4H SiC samples having $N_d - N_a \approx 5 \times 10^{19}$ cm$^{-3}$ (Fig. 2c) at $T$ = 10 K we observe the similar broad intense S-line having $S$ = 1/2, $g_{\parallel}$ = 2.0046(3), $g_{\perp}$ = 2.0010(3).

TABLE I. The parameters of spin Hamiltonian for N donors ($I$ = 1) in monocrystalline 4H SiC with various N donor concentrations.

| $N_d - N_a$, cm$^{-3}$ | Temp. range | Center | S | $g_{\perp}$ | $g_{\parallel}$ | $A_{\perp}$, MHz | $A_{\parallel}$, MHz | $D$, MHz | Ref. |
|---|---|---|---|---|---|---|---|---|---|
| $10^{17}$ | ≤ 40 K | $N_h$ | 1/2 | 2.0005(3) | 2.0063(3) | 2.7 | 2.9 | - | This work, Refs. 29-31 |
| | ≤ 40 K | $N_x$ | 1 | 2.0010(3) | 2.0053(3) | 25.6 | 25.6 | 1.7-14 | This work, Refs. 29-31 |
| | ≤ 60 K | $N_k$ | 1/2 | 2.0013(3) | 2.0043(3) | 51 | 50.95 | - | This work, Ref. 51 |
| $5 \times 10^{18}$ | ≤ 40 K | $N_k$ | 1/2 | 2.0013(3) | 2.0043(3) | 51 | 50.95 | - | This work, Ref. 51 |

Fig. 3 represents the EPR spectra temperature variation measured in monocrystalline 4H SiC with various N donor concentrations recorded in the broad temperature interval normalized to its maximum intensity values.

In 4H SiC samples having $N_d - N_a \approx 10^{17}$ cm$^{-3}$ at $T > 90$ K, no EPR spectra were observed (Fig. 3a). As is seen from Fig. 3a, in these samples, the single slightly asymmetric EPR S-line at $T \leq 90$ K was detected. Along with a single S-line, the $N_k$ triplet appeared at $T \leq 60$ K. With lowering the temperature to 40 K, the S-line vanishes, whereas the $N_h$ and $N_x$ triplets emerge in EPR spectra.

In 4H SiC having $N_d - N_a \approx 5 \times 10^{18}$ cm$^{-3}$, no EPR spectra were detected at $T > 190$ K, while a single asymmetric S-line was detected at $T < 190$ K, whereby the residual parts of the $N_k$ triplet appeared at $T < 20$ K (Fig. 3b). In 4H SiC having $N_d - N_a \approx 5 \times 10^{19}$ cm$^{-3}$ no EPR spectra were detected at $T > 150$ K, whereas a single asymmetric S-line was observed at $T \leq 150$ K (Fig. 3c).

The temperature variation FS ESE spectra recorded in monocrystalline 4H SiC having $N_d - N_a \approx 10^{17}$ cm$^{-3}$ at $T = 6-50$ K is shown in Fig. 4. At $T < 25$ K, $N_k$, $N_h$, and $N_x$ triplets are observed in FS ESE spectra, while at $T > 25$ K, the $N_k$ triplet is only detected, and no single S-line at $T = 40-50$ K in contrast with continuous wave EPR spectra was observed. Thus, we can conclude that the paramagnetic center that causes the emergence of the single S-line in the EPR spectrum has short relaxation times and, therefore, cannot be detected in FS ESE spectra.

The double integral intensity temperature variation of the S-line (that is proportional to the spin susceptibility, $\chi_{EPR}$) was obtained at $T = 5-150$ K in monocrystalline 4H SiC samples having $N_d - N_a \approx 5 \times 10^{19}$ cm$^{-3}$ (Fig. 5).

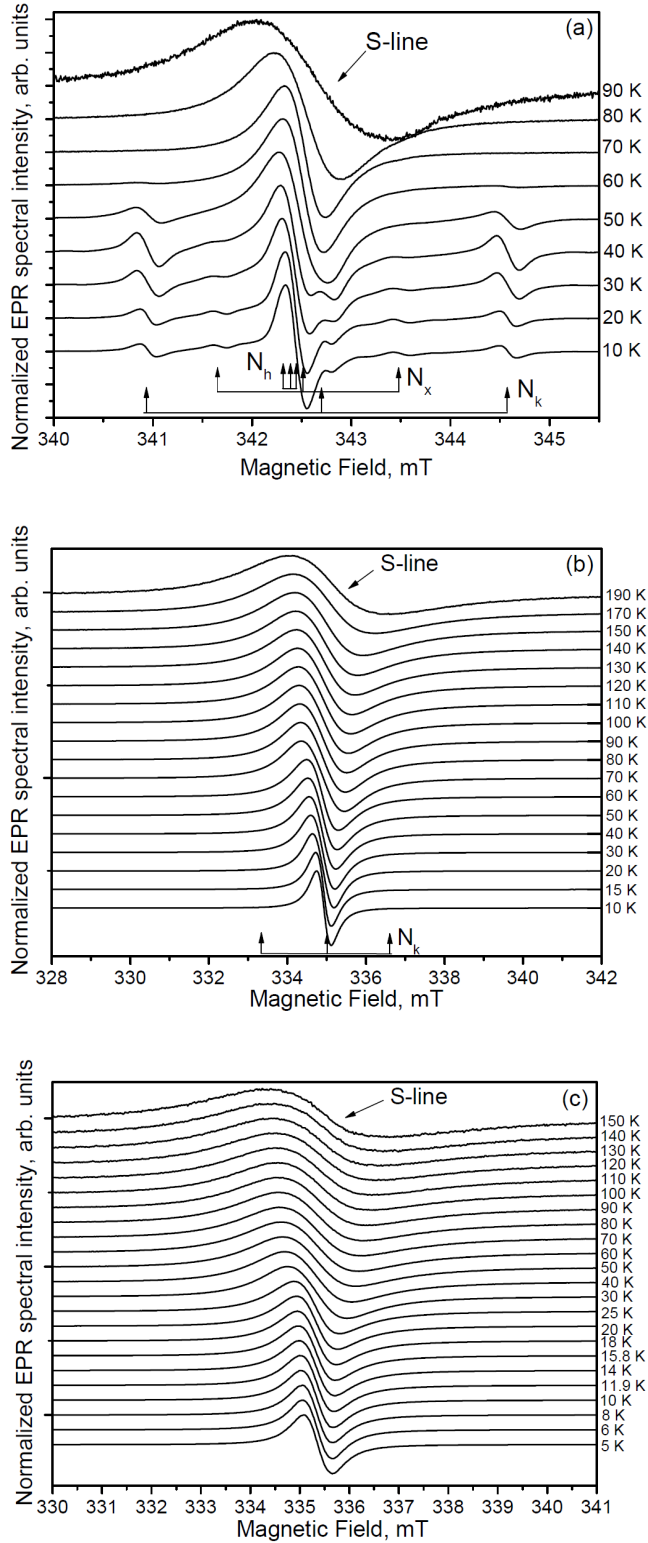

FIG. 3 Temperature variation of EPR spectra recorded in monocrystalline 4H SiC. (a) – $N_d - N_a \approx 10^{17}$ cm$^{-3}$, (b) – $N_d - N_a \approx 5 \times 10^{18}$ cm$^{-3}$, (c) – $N_d - N_a \approx 5 \times 10^{19}$ cm$^{-3}$. **B**||**c**, $T = 10$ K. The intensity of each EPR spectrum was normalized to its maximum value.

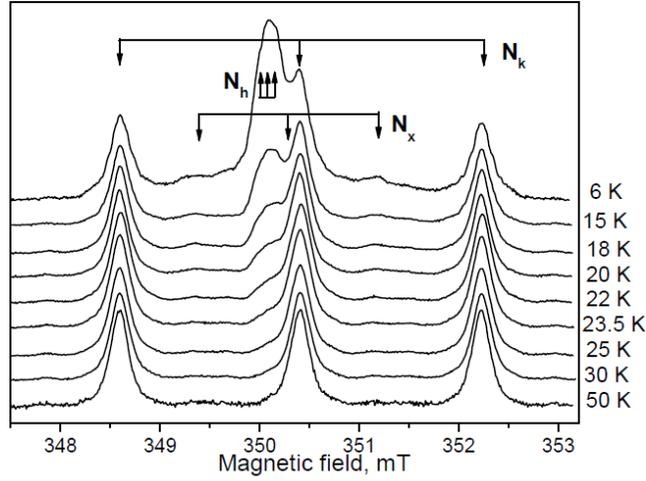

FIG. 4. Temperature variation of FS ESE spectra recorded in monocrystalline 4H SiC having $N_d - N_a \approx 10^{17}$ cm$^{-3}$ at **B**||**c**.

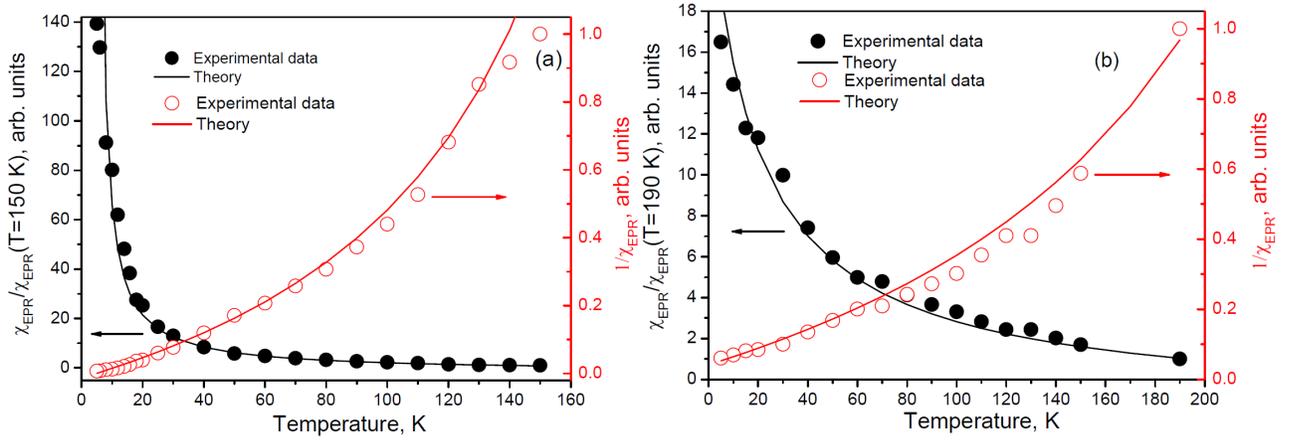

FIG. 5. Temperature variation of experimental $\chi_{EPR}$ (filled black circles) and $1/\chi_{EPR}$ (open red circles) values and corresponding fitting with Eq. (6) (solid black lines) and Eq. (7) (solid red lines) for S-line in 4H SiC monocrystals. (a) – $N_d - N_a \approx 5 \times 10^{19}$ cm$^{-3}$, (b) – $N_d - N_a \approx 5 \times 10^{18}$ cm$^{-3}$. **B** $\perp$ **c**.

Before double integration, the EPR spectra intensity at each temperature point was corrected to cavity $Q$-factor changes due to thermal losses of EPR cavity $Q$-factor. Afterward, the $\chi_{EPR}$ values were normalized to their minimum value. The temperature variation of $1/\chi_{EPR}$ revealed a linear behavior at low temperatures related to localized electrons (Curie-Weiss law), whereas, at higher temperatures, there is considerable curvature owing to the contribution of temperature-independent Pauli-like term $\chi_0$ to $\chi_{EPR}$ occurred due to the appearance of the non-localized conduction electrons.

According to Ref. 52, the theoretical description of the experimental data represented in Fig. 5 was done utilizing the following expressions:

$$\chi_{EPR}(T) = C/(T-\theta) + \chi_0, \qquad (6)$$

$$1/\chi_{EPR}(T) = (T-\theta)/(\chi_0(T-\theta) + C), \qquad (7)$$

where $C$ – Curie constant, $\theta$ – is the Curie-Weiss temperature.

Based on the fitting of Eq. (6) and Eq. (7) with experimental data shown in Fig. 5 for monocrystalline 4H SiC with $N_d - N_a \approx 5\times10^{19}$ cm$^{-3}$, we have obtained $\theta = 2.4$ K, showing that the weak ferromagnetic coupling is presented in the spin system. At the same time, in monocrystalline 4H SiC with $N_d - N_a \approx 5\times10^{18}$ cm$^{-3}$, the fitting gave the value of $\theta = -19.8$ K, which means a strong antiferromagnetic interaction exists in the spin system.

The asymmetry of the S-line EPR spectral shape of 4H SiC monocrystals occurs because the thickness of the skin layer is the same or less than the sample size caused by the increase in conductivity and, consequently, the diffusion time of the carriers throughout the skin layer is significantly shorter than the spin relaxation time. Consequently, the asymmetric Dysonian lineshape appears in the EPR spectra.[53,54] Fig. 6 shows an example of fitting Eq. (2) with experimental EPR spectra measured in monocrystalline 4H SiC having various N donor concentrations using different asymmetry ratio values. We have fitted Eq. (2) with experimental EPR spectra measured in monocrystalline 4H SiC having various N donor concentrations in a wide temperature interval at two different magnetic field orientations. The analysis of obtained data allowed us to assume that for the S-line, there is a temperature dependence of EPR linewidth, asymmetry ratio, and resonance magnetic field position.

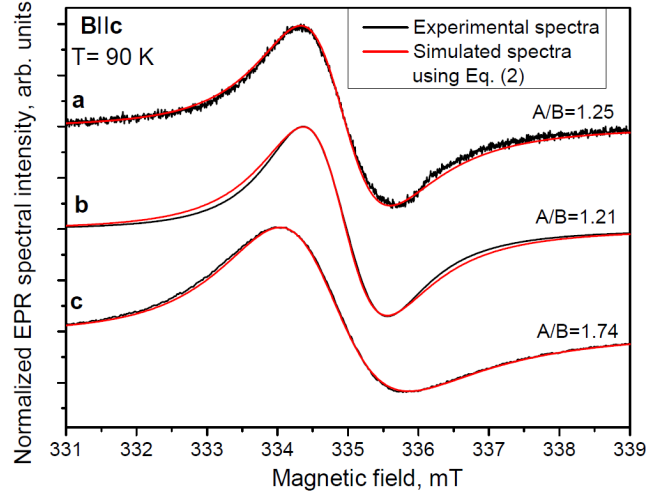

FIG. 6. The experimental EPR spectra S-line (solid black lines) and fitting of Dysonian lineshape using Eq. (2) (solid red lines) for monocrystalline 4H SiC. (a) – $N_d - N_a \approx 10^{17}$ cm$^{-3}$, (b) – $N_d - N_a \approx 5\times10^{18}$ cm$^{-3}$, (c) – $N_d - N_a \approx 5\times10^{19}$ cm$^{-3}$. $T = 90$ K, $\mathbf{B}\|\mathbf{c}$.

Fig. 7 represents the temperature variation of the Dysonian asymmetry ratio for the S-line in monocrystalline 4H SiC with various N donor concentrations. The *A/B* ratio for 4H SiC with $N_d - N_a \approx 10^{17}$ cm$^{-3}$ varies from 1.25 at 90 K to 1.0 (Lorentzian shape) at $T = 60$ K. For 4H SiC with $N_d - N_a \approx 5\times10^{18}$ cm$^{-3}$, the *A/B* ratio decreases gradually from 1.87 to 1.0 with the temperature from 190 K to 5 K. The *A/B* ratio temperature variation for 4H SiC with $N_d - N_a \approx 5\times10^{19}$ cm$^{-3}$ increases from 1.68 to 1.84 as the temperature decreases from 150 K to 120 K and afterward drops off with the temperature reaching 1.04 value at 5 K.

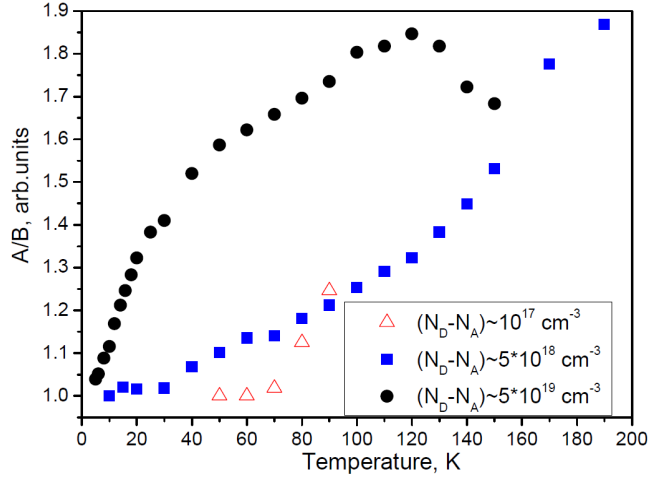

FIG. 7. Temperature variation of Dysonian asymmetry ratio for S-line as obtained from fitting of Dysonian lineshape using Eq. (2) in monocrystalline 4H SiC with $N_d - N_a \approx 10^{17}$ cm$^{-3}$ (open red triangles), $N_d - N_a \approx 5\times10^{18}$ cm$^{-3}$ (solid blue squares), $N_d - N_a \approx 5\times10^{19}$ cm$^{-3}$ (solid black circles) at $T = 90$ K, **B**‖**c**.

The magnetic resonance field position ($B_0$) variation with the temperature was derived from the Eq. (2) simulation with EPR spectra in monocrystalline 4H SiC having various N donor concentrations. Based on the obtained $B_0$ values, we have estimated the temperature variation of $g_\|$ and $g_\perp$ for the S-line using the expression: $g = h \times v_0 / \mu_B \times B_0$ (Fig. 8). In 4H SiC samples having $N_d - N_a \approx 10^{17}$ cm$^{-3}$ with the temperature decrease $g_\|$ shifts from 2.0053(3) to 2.0061(3) while $g_\perp$ shows a slight shift from 2.0010(3) to 2.0008(3) only. The 4H SiC having $N_d - N_a \approx 5\times10^{18}$ cm$^{-3}$ reveals the shift of $g_\|$ value from 2.0052(3) to 2.0045(3), and $g_\perp$ value varies in the limit of error from 2.0011(3) to 2.0010(3) with the temperature decrease from 190 K down to 5 K. And in 4H SiC samples with $N_d - N_a \approx 5\times10^{19}$ cm$^{-3}$ the decrease in temperature from 150 K down to 5 K leads to the decrease of $g_\|$ value from 2.0057(3) to 2.0044(3) and $g_\perp$ value from 2.0019(3) to 2.0011(3).

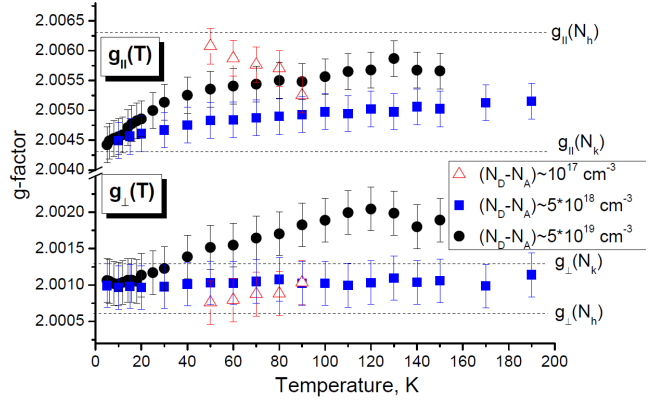

FIG. 8. Temperature dependence of $g_\parallel$ and $g_\perp$ for S-line derived from the simulation of EPR spectra using Eq. (2) in monocrystalline 4H SiC. Open red triangles – $N_d - N_a \approx 10^{17}$ cm$^{-3}$, solid blue squares – $N_d - N_a \approx 5\times10^{18}$ cm$^{-3}$, solid black circles – $N_d - N_a \approx 5\times10^{19}$ cm$^{-3}$. Dot lines show the position of $g_\parallel$ and $g_\perp$ for $N_k$ and $N_h$ centers.

Fig. 9 represents the EPR linewidth ($\Delta B_{pp}^L$) temperature dependence for the S-line derived from the simulation of the experimental spectra EPR recorded in monocrystalline 4H SiC with different N donor concentrations at $\mathbf{B} \perp \mathbf{c}$ with Eq. (2).

For the description of the experimental data in Fig. 9, the following expression similar to highly-doped 6H SiC can be used:[55]

$$\Delta B_{pp}(T) = \Delta B_0 + b \times T + c \frac{\Delta}{\exp(\Delta/T)-1} \quad (8)$$

where $\Delta B_0$ – is the residual EPR linewidth at $T = 0$ K, $b$ is characterized by the thermal fluctuations of the exchange coupling of magnetic moments of the localized electrons with the non-localized ones at low temperatures (Korringa relaxation), the third term is related to the localized electrons relaxation at higher temperatures through excited levels with the energy $\Delta$ related to the ground state and caused by the exchange coupling of localized electrons with the non-localized ones (Orbach process), and *c* determines the orbit-lattice interaction strength.

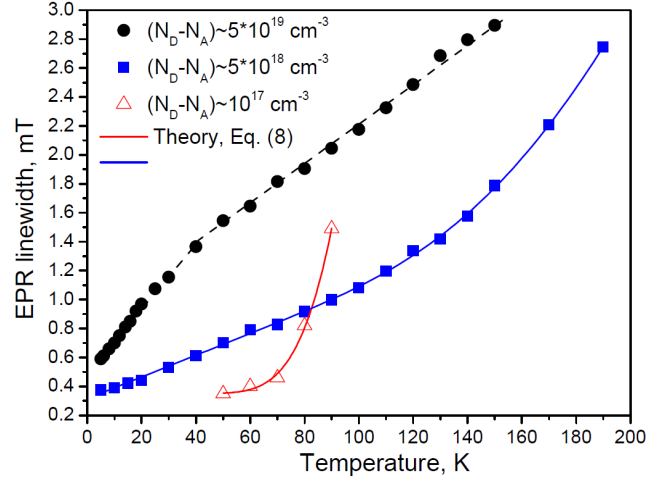

FIG. 9. Temperature variation of EPR linewidth ($\Delta B_{pp}^L$) for S-line derived from the simulation of EPR spectra using Eq. (2) in monocrystalline 4H SiC. Open red triangles – $N_d - N_a \approx 10^{17}$ cm$^{-3}$, solid blue squares – $N_d - N_a \approx 5 \times 10^{18}$ cm$^{-3}$, solid black circles – $N_d - N_a \approx 5 \times 10^{19}$ cm$^{-3}$. Red and blue solid lines show the fitting output using Eq. (8). The black dashed line guides the eye. **B**⊥**c**.

In 4H SiC having $N_d - N_a \approx 10^{17}$ cm$^{-3}$, only the first and third terms were considered for fitting experimental data with $\Delta \approx 57$ meV. In 4H SiC having $N_d - N_a \approx 5 \times 10^{18}$ cm$^{-3}$, all three terms from Eq. (8) were used to approximate experimental data with $\Delta \approx 68$ meV.

In the 4H SiC with $N_d - N_a \approx 5 \times 10^{19}$ cm$^{-3}$, the S-line EPR width increases linearly with the temperature with the slight change slope at ~ 40 K related to the second term in Eq. (8). This fact is consistent with the case when the spin-lattice relaxation of non-localized electron spins in metals is governed by modulation of spin-orbital interaction through lattice vibrations.[56]

**Discussion**

The temperature behavior of the spin coupling between localized and non-localized electrons in 4H SiC monocrystals with uncompensated N donor concentration from $10^{17}$ cm$^{-3}$ to $5\times10^{19}$ cm$^{-3}$ have been examined by EPR and MW cavity perturbation techniques at $T$ = 4.2-300 K.

Based on the study of the temperature variation of the conductivity, the activation energies $\varepsilon_1$, $\varepsilon_2$, and $\varepsilon_3$ have been estimated by fitting the temperature variation of conductivity with terms from Eq. (5). For 4H SiC having $N_d - N_a \approx 10^{17}$-$5\cdot10^{18}$ cm$^{-3}$ the obtained value $\varepsilon_1$ = 40 meV is in perfect agreement with a recently reported value of 40 meV for $N_h$ in 4H SiC wafers with $N_d - N_a \approx 10^{18} \div 5\times10^{18}$ cm$^{-3}$ in Ref. 47 while for 4H SiC with $N_d - N_a \approx 5\times10^{19}$ cm$^{-3}$ the obtained $\varepsilon_1$ value 120 meV well agrees with the energy level of the $N_k$ donors having 105-125 meV.[18,19]

We can explain the differences between obtained values of $\varepsilon_1$ in the investigated samples and the relationship between N donor concentration in the samples and activation energy.

In N-doped SiC, it is expected that the activation energy needed for an electron to be excited from the N donor level to the conduction band should decrease as N donor concentration increases, providing more available donors for electron donation to the conduction band, which makes it easier for electrons to be excited from the donor level. However, the relationship between N donor concentration and activation energy can become more complex on the metal side of the semiconductor-metal transition. In 4H SiC, the metal side of the semiconductor-metal transition, as follows from Mott criteria ($a_B N_{crit}^3 \approx 0.25$),[57] occurs at N donor concentration higher than critical carrier concentration $N_{crit} \sim 8\cdot10^{18}$ cm$^{-3}$. At such high N doping concentrations in SiC, a self-passivation effect due to the formation of N aggregates[58] or N-carbon complexes[59] can appear responsible for the low effectiveness of high-dose implantation of N and lower conductivity. Therefore in the monocrystalline 4H SiC with $N_d - N_a \approx 5\times10^{19}$ cm$^{-3}$, the self-passivation effect, formation of N aggregates or N-carbon complexes could be the reason for the increase of the $\varepsilon_1$ value and, as a result, the decrease of the conductivity.

Generally, the activation energy required for an electron to hop from occupied to unoccupied donors should decrease with increasing donor concentration in an n-type semiconductor. However, in our case, the $\varepsilon_3$ energy value was 5 meV in all investigated samples. The Hall measurements at low temperatures should be performed in 4H SiC monocrystals with $N > N_{crit}$ to explain this phenomenon in more detail.

As follows from Fig. 2a, the EPR spectrum in monocrystalline 4H SiC with $(N_D - N_A) \sim 10^{17}$ cm$^{-3}$ consists of three triplets: $N_k$, $N_h$, and $N_x$, in accordance with data represented in Refs. 29-31. In 4H SiC samples with $N_d - N_a \geq 5 \times 10^{18}$ cm$^{-3}$ at $T = 10$ K, we observe the broad intense S-line with spin Hamiltonian parameters close to those previously reported an intense single line in 4H SiC wafers with $N_d - N_a \approx 10^{18}$ cm$^{-3}$ at $T = 4.2$ K:[30] $g_\parallel = 2.0054(1)$, $g_\perp = 2.0011(3)$. The discrepancy in values obtained in our work and Ref. 30 can be related to lower N concentration in the samples measured in Ref. 30.

The analysis of temperature variation of S-line EPR spectra in monocrystalline 4H SiC with various N concentrations allowed us to establish that the S-line has temperature-dependent integral intensity, linewidth, line asymmetry, and resonance magnetic field position.

The temperature variation of the asymmetry ratio of S-line in monocrystalline 4H SiC having various N donor concentrations agreed with the conductivity temperature dependence as deduced from measurements by the MW cavity perturbation method. At high temperatures where the conductivity is high, the Dysonian lineshape was observed. In contrast, at low temperatures, the Lorentzian lineshape dominates in the EPR spectrum for the S-line due to low conductivity in the samples. The fact that temperature variation of the Dysonian asymmetry ratio for S-line in monocrystalline 4H SiC with $N_d - N_a \approx 5 \times 10^{19}$ cm$^{-3}$ slightly rises with decreasing the temperature from 150 K to 120 K and afterward drops off with the temperature, we can explain by the transition process of electrons from energy levels of N donors to the conduction band taking place at $T = 170\text{-}90$ K.

The temperature dependence of spin susceptibility was described by two contributions: Curie-Weiss behavior at low $T$ values due to localized electrons and Pauli-like behavior at higher temperatures owing to the existence of non-localized electrons.

The presence of the coupling of localized electrons and non-localized ones can be derived from the resonance magnetic field position temperature dependence (and thus the $g$-factor). From the high-temperature part of the $g$-factor temperature dependence where non-localized centers dominate in EPR spectra, one can suppose that the free electrons in 4H SiC monocrystals are characterized by $g_\parallel = 2.0053(3)$ $g_\perp = 2.0011(3)$ for $N_d - N_a \leq 5 \times 10^{18}$ cm$^{-3}$ and $g_\parallel = 2.0057(3)$ and $g_\perp = 2.0019(3)$ for $N_d - N_a \approx 5 \times 10^{19}$ cm$^{-3}$. At the same time, we have found that for localized electrons (dominated at low temperatures) in monocrystalline 4H SiC with $(N_D - N_A) \sim 10^{17}$ cm$^{-3}$ $g_\parallel = 2.0061(3)$ and $g_\perp = 2.0008(3)$ being close to $g$-values of N$_h$ donors in 4H SiC,[29-31] while in 4H SiC with $N_d - N_a \geq 5 \times 10^{18}$ cm$^{-3}$, the localized electrons are characterized by $g_\parallel = 2.0044(3)$ and $g_\perp = 2.0011(3)$ that are close to $g$-values of N$_k$ donors in 4H SiC.[51] Thus, we can conclude that N$_h$ donors play the role of localized centers in 4H SiC with $N_d - N_a \approx 10^{17}$ cm$^{-3}$, whereas, in 4H SiC with $N_d - N_a \geq 5 \times 10^{18}$ cm$^{-3}$, the localized centers should be related to N$_k$ donors.

The corresponding $g_\parallel$ and $g_\perp$ values for localized electrons and non-localized ones obtained in this work and the literature data are given in Table II. In addition, in Table II, we show the average value of $g$-factor: $g_{av} = (2g_\perp + g_\parallel)/3$ for non-localized electrons that turned out in monocrystalline 4H SiC with $N_d - N_a \approx 10^{17} \div 5 \times 10^{18}$ cm$^{-3}$ to be close to the $g$-factor of a free electron, while for $N_d - N_a \approx 5 \times 10^{19}$ cm$^{-3}$ it has a slightly higher value.

The short spin relaxation times for the S-line, as deduced from temperature-dependent FS ESE measurements, also support the exchange nature of this paramagnetic center.

The temperature dependence of S-line EPR linewidth monocrystalline 4H SiC having $N_d - N_a \approx 10^{17}$ cm$^{-3}$ was characterized by the Orbach process, whereas for $N_d - N_a \approx 5\times10^{18}$ cm$^{-3}$, it is governed both by Korringa and Orbach relaxation and for $N_d - N_a \approx 5\times10^{19}$ cm$^{-3}$ (on the metallic side of metal-semiconductor transition) it is linear with a slight slope corresponding to Korringa relaxation process only.

TABLE II. Spin Hamiltonian parameters of localized and non-localized electrons monocrystalline 4H SiC with various N donor concentrations as deduced from the $g$-factor temperature dependence for S-line represented in Fig. 8.

| $N_d - N_a$, cm$^{-3}$ | Localized electrons (low temperatures) | | Non-localized electrons (high temperatures) | | | Ref. |
|---|---|---|---|---|---|---|
| | $g_\perp$ | $g_\parallel$ | $g_\perp$ | $g_\parallel$ | $g_{av}$ | |
| $10^{17}$ | 2.0008(3) | 2.0061(3) | 2.0010(3) | 2.0053(3) | 2.0024(3) | This work |
| $10^{18}$ | 2.0011(3) | 2.0054(1) | - | - | - | Ref. 30 |
| $5\times10^{18}$ | 2.0010(3) | 2.0045(3) | 2.0011(3) | 2.0052(3) | 2.0025(3) | This work |
| $10^{19}$ | 2.0003(3) | 2.0045(3) | - | - | - | Ref. 32 |
| $5\times10^{19}$ | 2.0011(3) | 2.0044(3) | 2.0019(3) | 2.0057(3) | 2.0032(3) | This work |

It is known that for N donors in 4H SiC, the valley-orbit splitting ($\Delta E_{v.-o.}$) values between the ground 1S(A$_1$) and excited 1S(E) energy levels are $\Delta E_{v.-o.}$ = 45.5 meV for N$_k$ in Ref. 60, 61 and $\Delta E_{v.-o.}$ = 7.0-7.6 meV for N$_h$ in 4H SiC.[61-66] Thus, the $\Delta$ values (57-65 meV) derived from the fitting procedure of Eq. (8) with the temperature variation of the S-line linewidth can correspond to $\Delta E_{\text{v.-o.}}$ values of $N_k$ donors in monocrystalline 4H SiC with $N_d - N_a \leq 5 \times 10^{18}$ cm$^{-3}$.

**Conclusions**

The magnetic and electronic properties of 4H SiC monocrystals having N donor concentrations from $10^{17}$ cm$^{-3}$ to $5 \times 10^{19}$ cm$^{-3}$ grown by sublimation sandwich and modified Lely methods were studied by X-band EPR and cavity perturbation method in a broad temperature range.

The temperature dependence of the MW conductivity is characterized by following processes in monocrystalline 4H SiC: conduction electrons scattering via ionized N donors at high temperatures for $N_d - N_a \geq 5 \times 10^{18}$ cm$^{-3}$, thermal activation of carriers density in the conduction band with an activation energy of $\varepsilon_1$ equal to 40 meV for monocrystalline 4H SiC with $N_d - N_a \approx 10^{17} \div 5 \times 10^{18}$ cm$^{-3}$ and 120 meV for 4H SiC samples $N_d - N_a \approx 5 \times 10^{19}$ cm$^{-3}$ and nearest-neighbor hopping processes of electrons from neutral N donors to empty positively charged N donors with activation energy $\varepsilon_3$ of 5 meV for $N_d - N_a \geq 5 \times 10^{18}$ cm$^{-3}$.

A single S-line ($S = 1/2$) of Dysonian lineshape governs the EPR spectra of 4H SiC monocrystals at high temperatures. The analysis of FS ESE measurements, along with the temperature variation of the integral intensity, line asymmetry, and resonance magnetic field position of this S-line in monocrystalline 4H SiC with various N donor concentrations, enabled us to conclude that this paramagnetic center is due to exchange coupling among localized electrons and non-localized ones. The role of localized centers in monocrystalline 4H SiC with $N_d - N_a \approx 10^{17}$ cm$^{-3}$ play the $N_h$ donors, while in 4H SiC monocrystals with $\geq 5 \times 10^{18}$ cm$^{-3}$, the localized centers are related to $N_k$.

Based on the analysis of S-line EPR linewidth temperature variation in 4H SiC with $N_d - N_a \leq 5\times10^{18}$ cm$^{-3}$, the splitting between the ground 1S(A$_1$) and excited 1S(E) energy levels of 57-65 meV was derived that corresponds to valley-orbit splitting value of N$_k$ donors in 4H SiC.

**Acknowledgments**

The authors are grateful to Dr. S. Friedländer for the temperature measurements of 4H SiC samples with $(N_D - N_A) \sim 10^{17}$ cm$^{-3}$. D. Savchenko and J. Lančok are grateful for support from Operational Program Research, Development and Education financed by European Structural and Investment Funds and the Czech Ministry of Education, Youth and Sports [Project SOLID21 CZ.02.1.01/0.0/0.0/16 019/0000760].